\documentclass[a4paper,amsmath,amssymb,amsfonts,superscriptaddress,showpacs,twocolumn]{revtex4-2}
\usepackage{graphicx}
\usepackage{braket}
\usepackage{hyperref}

\newcommand{\be}{\begin{equation*}}
\newcommand{\ee}{\end{equation*}}
\newcommand{\bea}{\begin{eqnarray*}}
\newcommand{\eea}{\end{eqnarray*}}
\begin{document}

\title{Variational optimization of continuous matrix product states}
\author{Beno\^it~\surname{Tuybens}}
\author{Jacopo~\surname{De Nardis}}
\author{Jutho~\surname{Haegeman}}
\author{Frank~\surname{Verstraete}}
\affiliation{Department of Physics and Astronomy, Ghent University, Krijgslaan 281, S9, B-9000 Ghent, Belgium}

\begin{abstract}
Just as matrix product states represent ground states of one-dimensional quantum spin systems faithfully, continuous matrix product states (cMPS) provide faithful representations of the vacuum of interacting field theories in one spatial dimension. Unlike the quantum spin case however, for which the density matrix renormalization group and related matrix product state algorithms provide robust algorithms for optimizing the variational states, the optimization of cMPS for systems with inhomogeneous external potentials has been problematic. We resolve this problem by constructing a piecewise linear parameterization of the underlying matrix-valued functions, which enables the calculation of the exact reduced density matrices everywhere in the system by high-order Taylor expansions. This turns the variational cMPS problem into a variational algorithm from which both the energy and its backwards derivative can be calculated exactly and at a cost that scales as the cube of the bond dimension. We illustrate this by finding ground states of interacting bosons in external potentials, and by calculating boundary or Casimir energy corrections of continuous many-body systems with open boundary conditions. 
\end{abstract}
%\pacs{03.67.-a,64.60.ae,11.25.Hf}
\maketitle

\noindent The density matrix renormalization group (DMRG) and variational matrix product state algorithms have revolutionized the understanding of static and dynamic properties of quantum spin systems \cite{White}. Many of the low dimensional strongly interacting systems of current interest however, such as the ones realized in optical lattices \cite{Bloch} and atom chips \cite{Schmiedmayer,Erne2018,Bouchoule}, are not described by lattices but rather by continuum field theories \cite{Cazalilla,Seiringer2008}. This led to the introduction of continuous matrix product states (cMPS) \cite{Verstraete,haegeman2013calculus}, the natural continuum analogue of MPS, which can also be understood as a matrix-valued generalization of coherent states used in describing Bose-Einstein condensates. cMPS have been explored in the context of relativistic systems \cite{haegeman2010applying}, Bose gasses with long-range interactions \cite{rincon2015lieb}, atomtronics \cite{draxler2017continuous} and various applications involving coupled fields \cite{quijandria2014continuous,quijandria2015continuous,chung2016new}. The cMPS construction is closely related to the output fields in a setup of cavity quantum electrodynamics, and thus also holds potential for experimental realization  \cite{osborne2010holographic,enciso2013cavity,eichler2015exploring}.

The cMPS ansatz is a functional of matrix-valued functions $Q(x)$, $R_\alpha(x)$ (the variational parameters), given by
\[ \ket{\psi[Q,R_\alpha]}=\bra{0}\mathcal{P}\mathrm{e}^{\int_{0}^{L}\mathrm{d}x\, Q(x) \otimes \mathsf{1} +\sum_\alpha R_{\alpha}(x) \otimes\psi_{\alpha}^\dagger(x)}\ket{B}\ket{\Omega}\]
where $\psi_\alpha^\dagger(x)$ are field operators in second quantization satisfying the usual commutation or anti-commutation relations and $\alpha$ labels the different modes or particle species in the system. $\bra{0}$ and $\ket{B}$ are boundary vectors on the $D$-dimensional virtual ancilla space (where $Q(x)$, $R_\alpha(x)$ live), and $\ket{\Omega}$ is the empty vacuum state satisfying  $\psi_\alpha(x)\ket{\Omega}=0$. When applying the time dependent variational principle (TDVP) to the cMPS manifold, a non-trivial matrix generalization of the Gross-Pitaevskii equation, known as the ``quantum Gross Pitaevskii equation'', is obtained \cite{Haegeman}. It has however turned out to be very challenging to integrate this matrix-valued partial differential equation, in particular for finite systems, due to the existence of divergencies occurring when inverting the left and/or right reduced density matrices. So far, cMPS algorithms have been restricted to ground state calculations in infinite systems with constant \cite{ganahl2017continuous} or periodic potentials \cite{Ganahl1}.

In this Letter, we propose a cMPS groundstate algorithm for finite systems with open boundary conditions, using a piecewise linear ansatz for the matrix-valued functions $Q(x)$ and $R(x)$. The local reduced density matrices on the correspondig segments are modelled using a quickly converging Taylor series expansion. In comparison to the more sophisticated basis spline approach from Ref.~\onlinecite{Ganahl1}, our approach has the crucial advantage that both the energy and its backwards derivative can be calculated exactly and at a very small cost, without having to refer to a lattice discretization and matrix product state techniques.

In more detail, the ansatz consists of specifying a mesh $\{x_k\}$ of strictly increasing coordinates,  $D\times D$ matrices $Q_k=Q(x_k)$, $R_k=R(x_k)$, and defining $Q(x)$, $R(x)$ on all other points by linear interpolation:
\begin{eqnarray*} %Q(x_k\leq x\leq x_{k+1})&=&  Q_k+\frac{x-x_k}{x_{k+1}-x_k} \left(Q_{k+1}-Q_k\right)\\
R(x_k\leq x\leq x_{k+1})&=&  R_k+\frac{x-x_k}{x_{k+1}-x_k} \left(R_{k+1}-R_k\right)
\end{eqnarray*}
and similar for $Q(x_k\leq x\leq x_{k+1})$. Using the language of finite element analysis, we are modelling $Q$ and $R$ using \emph{tent} functions. The left and right density matrices $\rho$ and $\sigma$, which encode the entanglement degrees of freedom, are defined by the Lindblad-like equations \cite{Verstraete,haegeman2013calculus}
\bea
\frac{\mathrm{d}\ }{\mathrm{d} x}\rho(x)&=& Q(x)^\dagger \rho(x)+\rho(x)Q(x)+R^\dagger(x)\rho(x)R(x)\\
-\frac{\mathrm{d}\ }{\mathrm{d} x}\sigma(x)&=& Q(x) \sigma(x)+\sigma(x)Q(x^\dagger)+R(x)\sigma(x)R(x)^\dagger(x)
\eea
with boundary conditions $\rho(0)=\ket{0}\bra{0}$ and $\sigma(L)=\ket{B}\bra{B}$. The key technical tool that makes the method successful, is formulating these reduced density matrices on the same segments as a Taylor series expansion
\bea \rho(x_k\leq x\leq x_{k+1})&=&\rho_k^{(0)}+\sum_{n=1}^{n_k} \left(\frac{x-x_k}{x_{k+1}-x_k}\right)^n \rho_k^{(n)}\\
 \sigma(x_k\leq x\leq x_{k+1})&=&\sigma_{k+1}^{(0)}+\sum_{m=1}^{m_k} \left(\frac{x_{k+1}-x}{x_{k+1}-x_k}\right)^m \sigma_{k+1}^{(m)}
 \eea
By plugging this ansatz into the Lindblad equations, we obtain a recursive triangular linear set of equations completely determining the matrices $\rho_k^{(n)}$ as a function of $Q_k,Q_{k+1}$, $R_k$, $R_{k+1}$ and $\rho_k^{(0)}$, and equally so for $\sigma_{k+1}^{(n)}$ as a function of  $Q_k$, $Q_{k+1}$, $R_k$, $R_{k+1}$ and $\sigma_{k+1}^{(0)}$. We also have the equations $\rho_{k+1}^{(0)}=\sum_{n=0}^{n_k}\rho_k^{(n)}$, $\sigma_{k}^{(0)}=\sum_{m=0}^{m_k}\sigma_{k+1}^{(m)}$. For a sufficiently fine mesh $\{x_k\}$, this recursion converges rapidly, such that the density matrices $\rho(x)$ and $\sigma(x)$ can be obtained with machine precision using only finite values of $m_k$ and $n_k$ (typical values are $n_k\simeq n_k\simeq 30$). We have indeed observed that it is instrumental to model the reduced density matrices $\rho(x)$ and $\sigma(x)$ up to a very high precision for the resulting algorithms to work robustly.

Once the reduced density matrices are determined, we can evaluate the energy. For simplicity, we consider a Lieb-Liniger Hamiltonian \cite{LiebLiniger} with kinetic energy term $(\mathrm{d} \psi_x^\dagger/\mathrm{d}x) (\mathrm{d} \psi_x/\mathrm{d}x)$, point interactions $g\psi_x^\dagger\psi_x^\dagger\psi_x\psi_x$, external potential $V(x)-\mu$, where the chemical potential $\mu$ indicates that we are working in the grand canonical ensemble and do not fix the number of particles. Finally, the Dirichlet boundary conditions $R(0)=R(L)=0$ (but free boundary conditions for $Q(0)$, $Q(L)$) corresponds to the boundary conditions imposed on particles in a box, where the wavefunction in first quantization vanishes at the boundaries. The energy is given by 
\bea E(Q,R,B)&=&\int_{0}^L \frac{  \bra{\rho(x)}H(x)\ket{\sigma(x)}}{\braket{\rho(0)|\sigma(0)}}\,\mathrm{d}x
\eea
with
\bea
H(x)&=&\mathcal{D} R(x) \otimes	 \overline{\mathcal{D} R(x)}+\left(V(x)-\mu\right)R(x)\otimes\overline{R(x)} \\
&&+g R(x)^2\otimes\overline{R(x)}^2
\eea
and
\bea
\mathcal{D}R(x) = [Q(x),R(x)]+\frac{\mathrm{d} R}{\mathrm{d} x}(x).
\eea
This integrand can readily be expressed in terms of the matrices $Q_k$, $R_k$, $\rho_k^{(n)}$, $\sigma_{k}^{(m)}$ and powers of $(x-x_k)/(x_{k+1}-x_k)$ and $(x_{k+1}-x)/(x_{k+1}-x_k)$. Hence, the integration can be carried out exactly in terms of the beta-functions $B(m+1,n+1)=\int_{0}^1x^m(1-x)^n\,\mathrm{d}x$. This allows for an \emph{exact} calculation of the energy for piecewise linear cMPS, as a sum 
\[\sum_k \sum_{n,m,a,b} B(n+a,m+b)\bra{\rho_k^{(n)}}\tilde{H}^{(a,b)}_k\ket{\sigma_{k+1}^{(m)}}\]
\noindent with $\tilde{H}^{(a,b)}_k$ the Taylor expansion coefficients of $H(x)$ in the interval $[x_k,x_{k+1})$. As is well known from the theory of finite elements, a piecewise linear approximation leads to a finite expectation value of the kinetic energy (Laplacian), as can readily be seen by doing partial integration.

The total computational complexity for computing the energy scales as $L \cdot D^3$, exactly as in the case of DMRG and variational MPS algorithms. Furthermore, the exact derivatives $\nabla_{Q_k} E$ and $\nabla_{R_k} E$ can also be calculated at effectively the same computational cost by making use of the idea of backwards differentiation. This is simple to implement as the determination of $\rho_k^{(n)}$ and $\sigma_k^{(m)}$ only involves simple matrix multiplications.  Consider the derivative of the energy towards the matrix elements $Q_k$ or $R_k$, we find
\bea \nabla_{k} E&=& E_{H}+E_L+E_R+E_N\\
E_{H}&=&\sum_{l=k-1}^{k}\sum_{n,m,a,b} B(n+a,m+b)\bra{\rho_l^{(n)}}\nabla_{k} \tilde{H}^{(a,b)}_l\ket{\sigma_l^{(m)}}\\
E_L&=&\sum_{l \geq k-1}\sum_{n,m,a,b} B(n+a,m+b)\bra{\nabla_{k} \rho_l^{(n)}}\tilde{H}^{(a,b)}_l\ket{\sigma_l^{(m)}}\\
E_R&=&\sum_{l \leq k}\sum_{n,m,a,b} B(n+a,m+b)\bra{\rho_l^{(n)}}\tilde{H}^{(a,b)}_l\ket{\nabla_{k}\sigma_l^{(m)}}\\
E_N&=&-E \frac{\left(\braket{\nabla_{k} \rho_k^{(0)}|\sigma_k^{(0)}}+\braket{\rho_k^{(0)}|\nabla_{k} \sigma_k^{(0)}}\right)}{\braket{\rho_k^{(0)}|\sigma_k^{(0)}}}
\eea
where $\nabla_k$ denotes either $\nabla_{Q_k}$ or $\nabla_{R_k}$. These terms are clearly reminiscent from the DMRG procedure, where the optimization of the energy at a site $k$ involves three different terms: the local energy term (corresponding to $E_{H}$) and the way this local tensor affects the energies at distant points ($E_L$ and $E_R$). Note however that the functional dependence for the cMPS case is a highly non-linear function of $Q_k$, as opposed to the linear dependence in the usual DMRG case. The key element in determining the Jacobians $\nabla_{k} \rho_l^{(n)}$ and $\nabla_{k} \sigma_l^{(m)}$ is the fact that all $\rho_{l}^{(n)}$ depend linearly on $\rho_{k+1}^{(0)}$, and furthermore that
\bea
\nabla_{Q_k} \rho_l^{(n)}=\frac{\partial \rho_l^{(n)}}{\partial \rho_l^{(0)}} \frac{\partial \rho_l^{(0)}}{\partial\rho_{l-1}^{(0)}}\cdots \frac{\partial\rho_{k+1}^{(0)}}{\partial Q_k}
\eea
and similar for $\nabla_{R_k}$. Given an operator $A$, we can hence efficiently calculate $\mathrm{tr}\left(A \frac{\mathrm{d} \rho_l^{(n)}}{\mathrm{d} Q_k}\right)=\braket{A|\frac{\mathrm{d} \rho_l^{(n)}}{\mathrm{d} Q_k}}$ by evolving $\bra{A}\rightarrow \bra{A'}=\bra{A}\frac{\mathrm{d}\rho_l^{(n)}}{\mathrm{d} \rho_l^{(0)}}\rightarrow \bra{A''}=\bra{A'}\frac{d \rho_l^{(0)}}{\mathrm{d}\rho_{l-1}^{(0)}}\rightarrow\cdots$. As $\rho_l^{(n)}$ is itself recursively defined as a function of $\rho_l^{(0)}$, this can easily be done by reversing the order of matrix multiplications done to produce the $\rho_l^{(n)}$. Indeed, this is a specific example of the general structure of reverse-mode differentiation as it appears in the context of differentiable programming. Note that it is important to do an auspicious bookkeeping to avoid double work. Effectively, what this approach is generating corresponds to the Taylor approximation of the integrated energy environments $\bra{H_L}(x)$ and $\ket{H_R}(x)$ defined by 
\bea
\frac{\mathrm{d}\ }{\mathrm{d} x} H_L(x) &=& Q(x)^\dagger H_L(x) + H_L(x) Q(x) + \\
&&R(x)^\dagger H_L(x) R(x) + H(x)
\eea
with boundary condition $H_L(0) = 0$, and similar for $H_R(x)$, the discrete analogue of which is also constructed in DMRG and related MPS algorithms.

We have thus demonstrated that the energy and its derivatives can be calculated at a cost that scales as $D^3$ in the bond dimension, linearly in the system size, and linearly in the number of Taylor expansion terms needed to capture the reduced density matrices up to machine precision. We can then use a standard non-linear optimization routine to determine the variational parameters $Q_k$ and $R_k$. There are two obvious choices: either we optimize over all parameters simultaneously, or sweep through the systems as in the case of DMRG and only optimize over the parameters corresponding to tensors in a certain localized region. In the latter case, the cost for calculating the derivatives does not depend anymore on the system size, given that necessary auxiliary quantities have correctly been archived. Unlike with DMRG, even the local problem is highly non-linear and thus not exactly solvable, so that in practice, the first seems preferable, as one can then immediately optimize global features. For the simulations in this manuscript, we used the global update scheme together with the standard L-BFGS algorithm \cite{nocedal1980updating}. It has proven to be very useful to initialize the cMPS with matrices $Q$, $R$ obtained from uniform cMPS calculations in the thermodynamic limit, brought into a gauge for which the left density matrix is equal to the right one, and to rescale the $R$ appropriately to zero around the boundary. 

A very desirably feature of this cMPS optimization method is that it is straightforward to refine the mesh over which we are linearly interpolating the matrices. It is certainly desirable to have a finer mesh around the boundaries. Also, it is possible to increase the bond dimension by embedding the $Q$ and $R$ matrices in larger matrices. An important note also concerns the gauge invariance of the cMPS that we use; for the most general cMPS, the resulting state is invariant under a local gauge transformation on the matrix-valued functions $Q(x)$ and $R(x)$. Because of the linear interpolation, this gauge freedom has disappeared and only one global gauge transform degree of freedom is retained. Without loss of generality, the right boundary vectors can hence be chosen to be a linear superposition $\alpha\ket{0}+\beta\ket{1}$, irrespective of the bond dimension $D$. For all simulations done in the Letter, we chose $\ket{B}=\ket{0}$.

Let us illustrate this cMPS algorithm with some examples, for which we again consider the interacting Bose gas using the aforementioned Lieb-Liniger Hamiltonian with an arbitrary external potential, open boundary conditions, in the grand canonical ensemble. We first set $V(x)=0$, $\mu=(4.75 \pi)^2$ and interaction term $g=10^6$, for which a quasi-exact solution is available. As the interaction is very strong, we are in the Tonks-Girardeau regime, and the energy can be computed in terms of free fermions filling the single particle modes: for the given value of $\mu$, the system will exhibit 4 fermions with a total energy  $(1+4+9+16)\pi^2 - 4\mu = -594.643$. From uniform cMPS calculations, it is well known that this Tonks-Girardeau limit is actually more demanding than the one with small effective $g/\rho$ (commonly denoted as $\gamma$), so this is a good test case. Running the variational cMPS algorithm with bond dimension $D=8$ and using $32$ equidistant interpolation points yields an energy equal to $E_{D=8} = -594.45$. The error is almost completely attributable to the linear interpolation. As shown in Fig.~1(a), the local density, the kinetic and the interaction terms are indistinguishable from the ones of the exact result.  We also calculated the entanglement spectrum as a function of position, see Fig.~1(b). Note that the exact ground state wavefunction with $N$ particles in the Tonks-Girardeau limit has an exact cMPS representation of bond dimension $2^N$, and is of the form $Q(x)=0$,
\[ R(x)=\sum_{k=1}^N Z^{\otimes (k-1)}\otimes \left[\begin{array}{cc} 0 & \sqrt{2}\sin(\pi k x)\\0 & 0\end{array}\right]\otimes \openone^{\otimes(N-k)}\]
and boundary conditions $\ket{B}=\ket{1}^{\otimes N}$.

\begin{figure}[t]
\begin{center}
		\includegraphics[width=\linewidth]{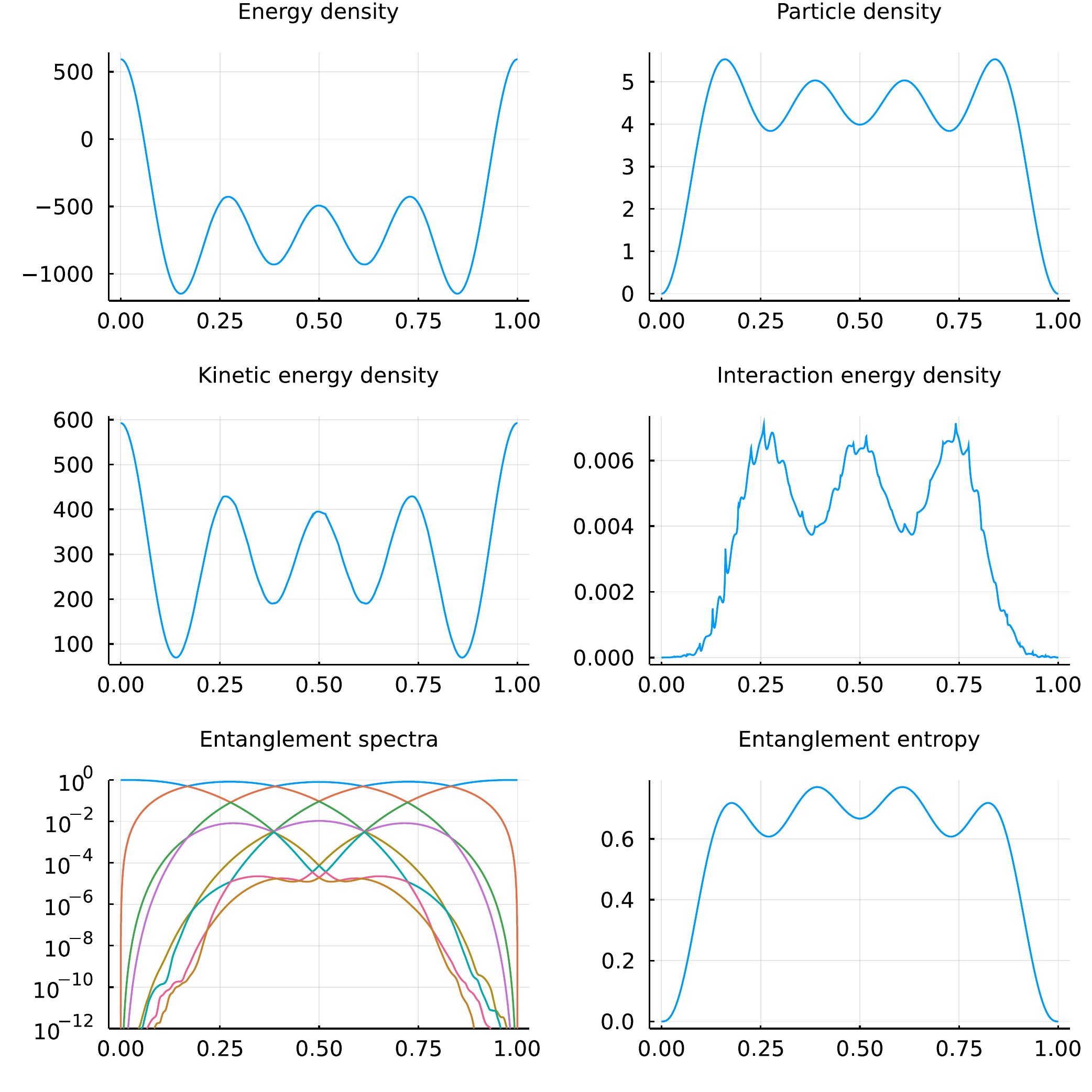}
\end{center}
\caption{Density and energy profiles of the ground state of an interacting Bose gas in a box of length $L=1$, $V(x)=0$, $\mu=(4.75\pi)^2$ and $g=10^6$, using cMPS with $D=8$ (top panels), and corresponding cMPS entanglement spectrum and entanglement entropy for a left-right bipartition as function of the cut position $x$ (bottom panel).}
\label{fig:MPOresults}
\end{figure}

\begin{figure}
	\begin{center}
		\includegraphics [width=\linewidth]{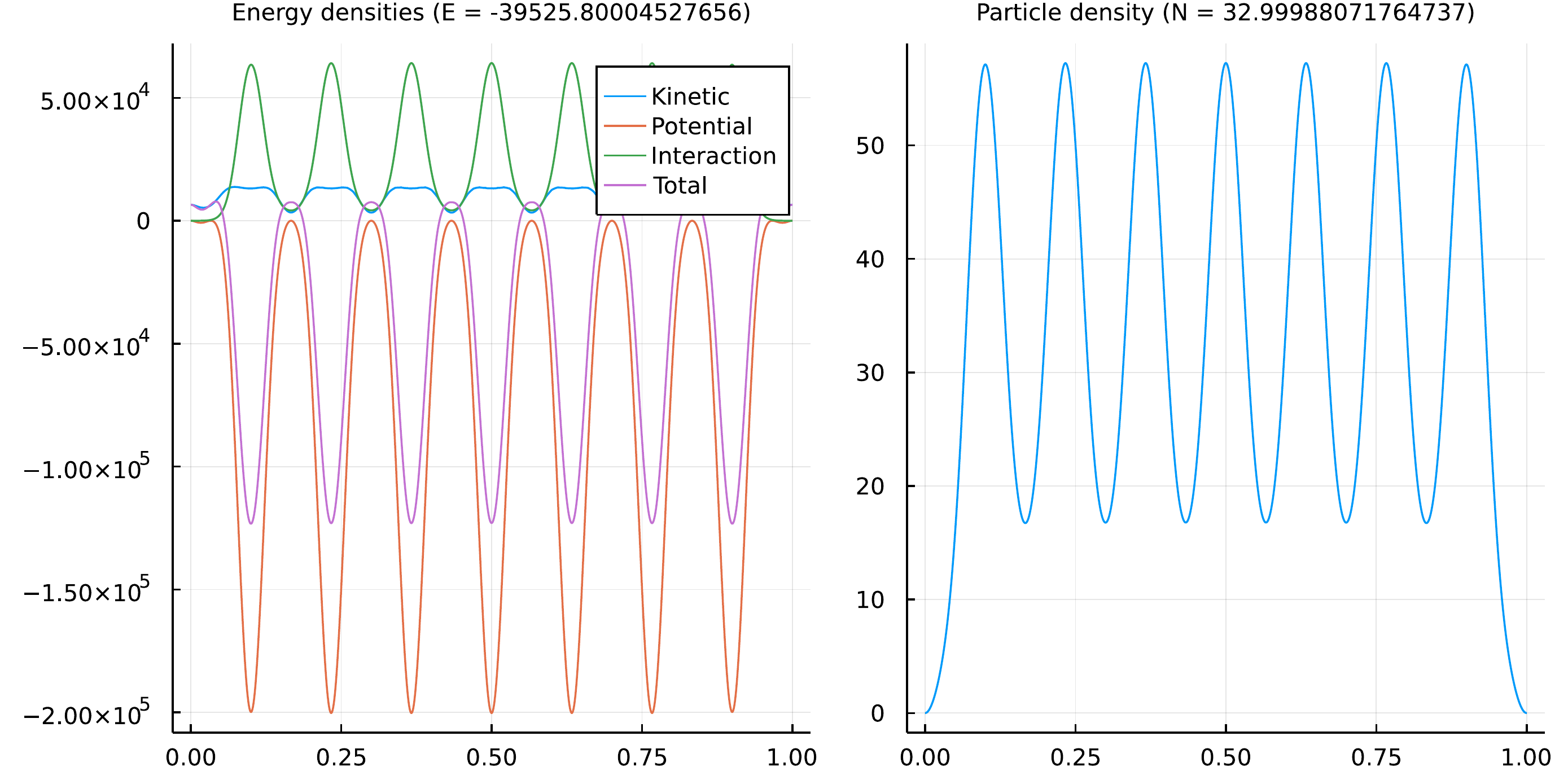} 
	\end{center}
	\caption{Density and energy profiles of the ground state of an interacting Bose gas in a box of length $L=1$, $V(x)=\mu \cdot \sin(15\pi x)$, $\mu=1749$ and $g=35$ (resulting in $\gamma=g/\rho=1.0509966$), using cMPS with $D=32$.}
	\label{LL_OBC}
\end{figure}

\begin{figure}
	\begin{center}
		\includegraphics [width=\linewidth ]{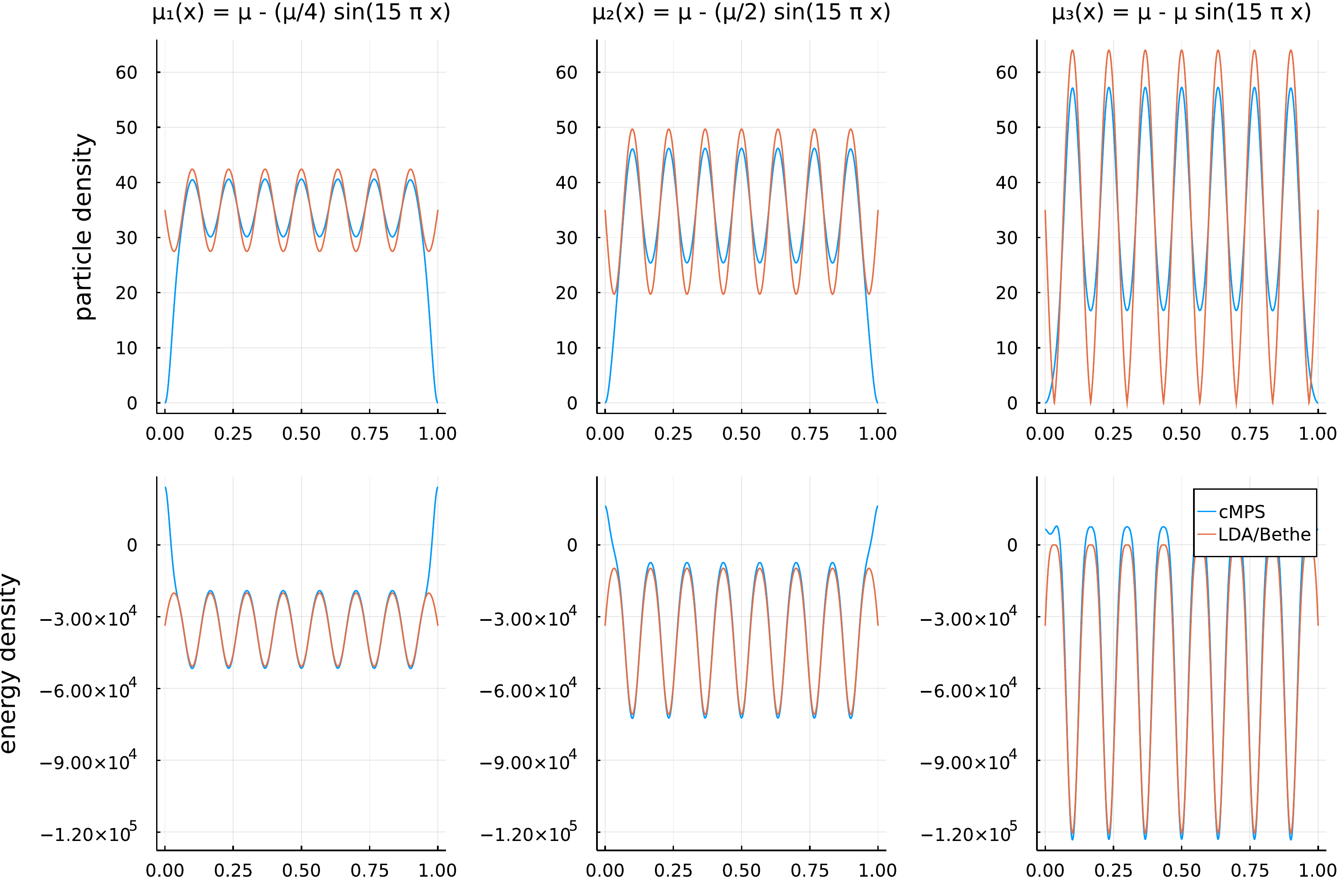} 
	\end{center}
	\caption{Density profiles corresponding to the parameters in Fig.~\ref{LL_OBC}, but with varying strength of the external potential. Comparison between $D=32$ cMPS results (blue line) and LDA approximation based on Yang-Yang thermodynamics of the Lieb-Liniger gas (red line).}		
	\label{LDA}
\end{figure}

Let us next move to a more challenging problem with many more particles. We consider a box with an external potential $V(x)=\mu \cdot \sin(15 \pi x)$, with $\mu=1749$ and  $g=35$. In that case, the dimensionless interaction strength is $g/\rho=1.0509966$, while the ground state density, kinetic energy, potential  energy and interaction energy profiles are depicted in Fig.~\ref{LL_OBC}. Note that there are around $33$ particles in the box, but that the cMPS ansatz does not exhibits particle number symmetry and hence allows for particle number fluctuations. We expect a sufficiently converged cMPS can take into account arbitrary external potentials, in contrast to the standard local density approximation (LDA) that is usually performed \cite{Amerongen,Bouchoule} by employing the  Yang-Yang thermodynamics \cite{Yang1969} of the Lieb-Liniger gas. In Fig. \ref{LDA} we show how the LDA can only predict the correct density profiles whenever $\partial_x \rho(x)/\rho(x) \ll 1 $, but it fails to reproduce the exact density profile obtained by cMPS outside this regimes, by modulating the strength of the external potential $V(x)$ in comparison to the chemical potential $\mu$. The effect on the energy density however is less significant.

%\onecolumngrid

%\twocolumngrid

\begin{figure}
	\begin{center}
		\includegraphics [width=0.9\linewidth]{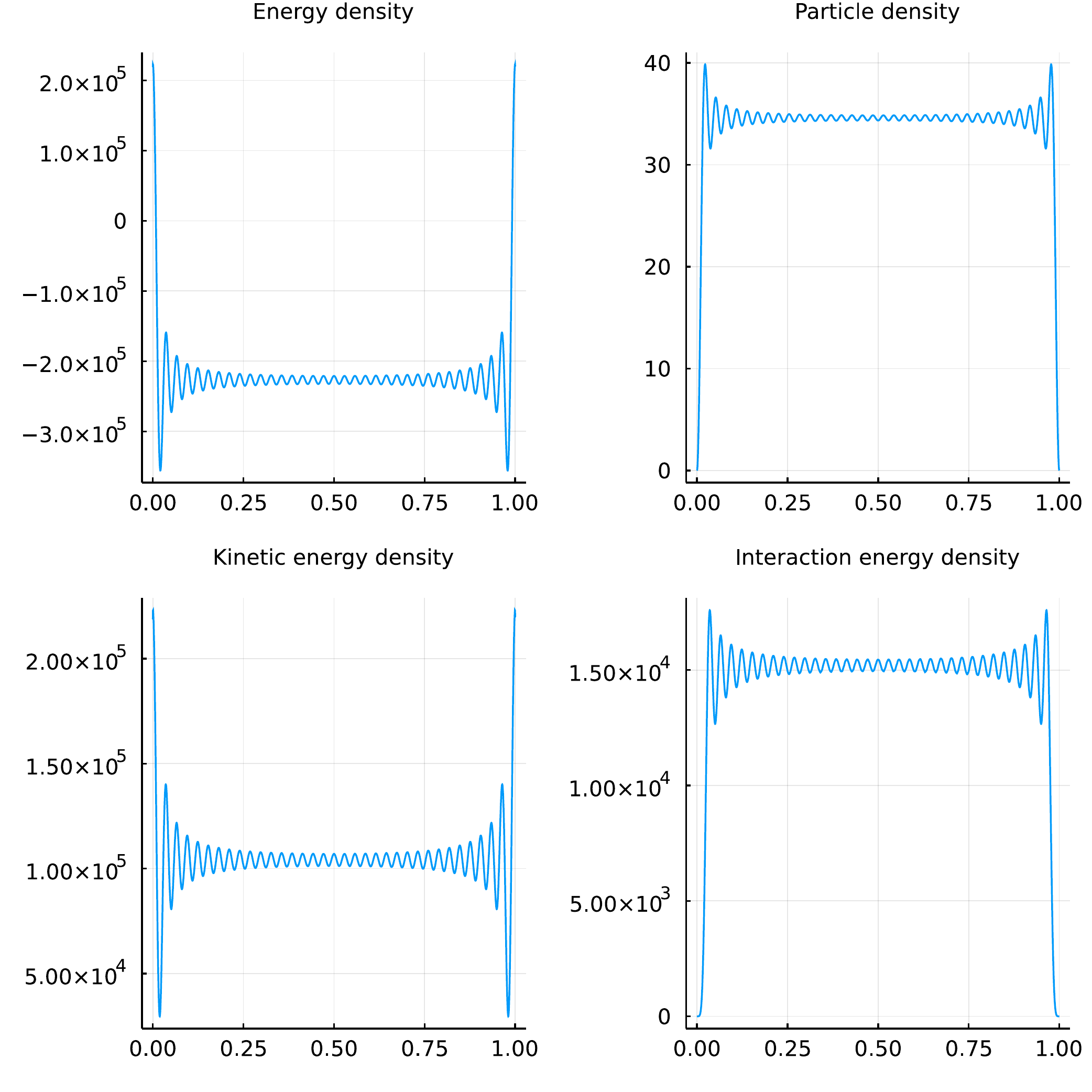} 
	\end{center}
		\caption{Density and energy profiles of the ground state of an interacting Bose gas in a box of length $L=1$, $V(x)=0$, $\mu=10000$ and $g=1000$ (resulting in $\gamma=29.288372$), using cMPS with $D=64$.}
	\label{LL_OBC1}
\end{figure}

\begin{figure}
	\begin{center}
		\includegraphics [width=\linewidth ]{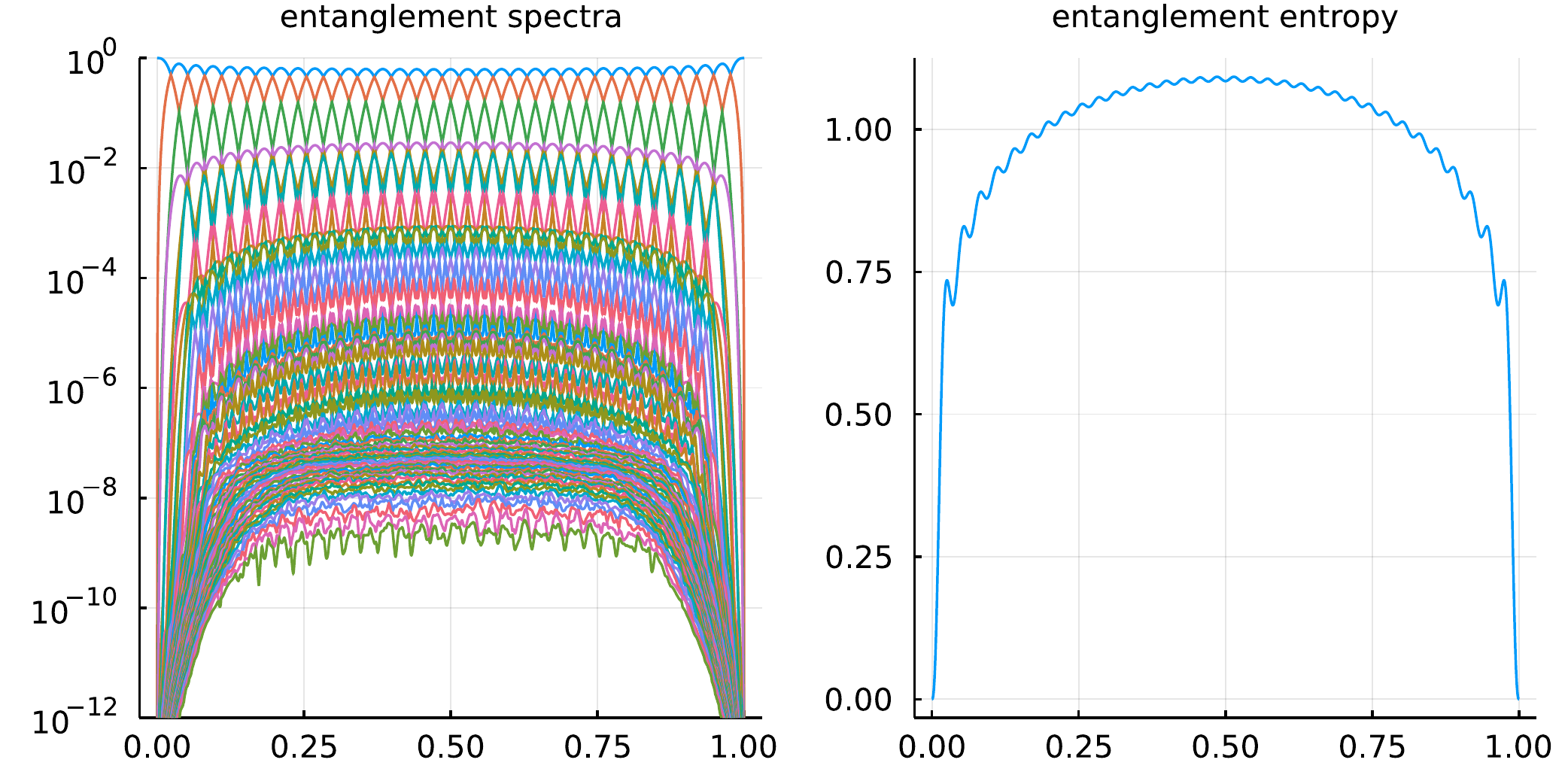} 
	\end{center}
	\caption{cMPS entanglement spectrum and entanglement entropy corresponding to the simulation in Fig.~\ref{LL_OBC1}, for a left-right bipartition at the 300 grid points on positions $x_k=(1-\cos(\pi k/300))/2$.}		
	\label{LL_OBC2}
\end{figure}

As a final example, we calculate the boundary or Casimir energy of such an interacting Bose gas. We therefore study the system in a box of length $L=1$ at $V(x)=0$, $\mu=10000$ and $g=1000$ using a cMPS with $D=64$ and $300$ grid points, and obtain the results in Fig.~\ref{LL_OBC1} and a total energy $E= -221056.1$, which includes the contribution of the chemical potential term. The total particle number in this system has expectation value $\braket{\hat{N}} = 33.9999$ with standard deviation $\sqrt{\braket{\hat{N}^2} - \braket{\hat{N}}^2} = 0.0256$. This has to be compared with the results from a uniform cMPS with bond dimension $D=64$ in the thermodynamic limit, which are $e_{\infty}=-226719.9$ for the energy density and $34.7840$ for the particle density. The resulting boundary or Casimir energy is given by $E_B=E - L e_{\infty} = 5663.8$. Note that the correct energy density obtained with the Bethe ansatz at  $\gamma = g/ \rho = 28.74888$ is given by $-226721.5$, whereas the Bethe ansatz prediction for the boundary energy (for open boundary conditions in the thermodynamic limit) is given by  $E_{B}=5676.0$ \cite{Gaudin,Reichert}, which is close to our value obtained for a system of length $L=1$. Note that the finite cMPS has very few particle fluctuations, and thus almost perfectly respects the $\mathsf{U}_1$ particle number symmetry. Indeed, the maximal value of the order parameter $|\braket{\psi(x)}|$ over all $x$ is given by $0.0016$ for the finite cMPS, whereas the uniform cMPS at the same bond dimension has $|\braket{\psi(x)}| = 1.5283$, indicating that the latter does significantly break the $\mathsf{U}_1$ in order to reduce entanglement. Related to this is the fact that the finite cMPS result captures the strong Friedel oscillations that exist in this system at these large values of $\gamma$, and that these  also persist in the entanglement spectrum (Fig.~\ref{LL_OBC2}).

%\twocolumngrid

In summary, this Letter introduces a scalable and robust variational method for studying interacting particles in a one-dimensional box using non-uniform continuous matrix product states, modelled using ideas from finite element analysis. A major technical bottleneck in cMPS algorithms was resolved by introducing a novel triangular set of coupled equations for the Taylor coefficients of the reduced density matrices, and this allowed for a quasi-exact determination of both the energy and its derivatives. There are many obvious ways in which these results can be extended. It is e.g.\ possible to use a quadratic interpolation scheme as opposed to a linear one; the triangular set of equations for the reduced density matrix is retained that way. It would also be possible to work in the left gauge for which $Q(x)=-\frac{1}{2}R(x)^\dagger R(x)+iK(x)$ and interpolate $K(x)$ instead. Alternatively, one can also investigate the use of spectral methods to model the cMPS matrix functions.

In a forthcoming publication, we will demonstrate how preconditioners can be constructed in order to speed up the optimization algorithms in terms of the local reduced density matrices, and how the time dependent variational principle (TDVP) can be adopted to the piecewise linear setting, so as to use non-uniform cMPS for time evolution while preserving local constants of motion such as the energy. We will also explore how cMPS algorithms can be used to simulate realistic higher-dimensional quantum many body systems, either ones that are confined in 2 directions such as in optical lattice experiments (for this, it will be enough to include multiple species $R_\alpha(x)$), or systems with a spherical symmetry, in which the cMPS would represent the radial part of the wavefunction (and angular components again appear as different modes).

The elephant in the room is the question whether cMPS algorithms of this kind exhibit any advantage over DMRG on a discretized grid. For relativistic quantum field theories with UV divergencies, where the lattice discretization plays the role of a very effective regulator, the answer is unclear. But for non-relativistic systems such as the ones created in optical lattices and atom chips, especially with shallow potentials, the lattice discretization seems to requires a very fine lattice spacing. The presence of multiple length scales then tends to result in convergence issues with DMRG \citep{Dolfi,Ganahl2}, making the approach particularly inconvenient at high densities. The cMPS approach in this Letter offers a cleaner and more natural approach to tackle such problems, and can more easily be combined with the iterative refinement strategies proposed to improve DMRG. 

\emph{Note:} The results in this paper can be obtained using the open-source Julia package CMPSKit.jl \footnote{B.~Tuybens, J.~Haegeman et al., CMPSKit.jl, \url{http://github.com/Jutho/CMPSKit.jl} (2021)} --- the details of which will be described elsewhere. A notebook with the necessary code to exactly reproduce the results of this Letter is also provided \footnote{Notebook arXiv:2006.01801 is hosted at \url{http://github.com/Jutho/cMPS-notebook} and can easily be viewed online at \url{http://nbviewer.jupyter.org/github/Jutho/cMPS-notebook/blob/master/arXiv\%3A2006.01801.ipynb}.}.

\begin{acknowledgements}
FV and JH acknowledge fruitful discussions with Lawrence Mitchell on the applicability of finite elements analysis to this problem, and discussions with Ignacio Cirac and Giacomo Giudice on a related project. This work has received funding from the European Research Council (ERC) under the European Union’s Horizon 2020 research and innovation programme (grant agreement No 715861 (ERQUAF) and 647905 (QUTE)). JDN is supported by a fellowship from the Research Foundation Flanders (FWO). BT is supported as an `FWO-aspirant' under contract number FWO17/ASP/199. 
\end{acknowledgements}

\bibliography{cmps_dmrg_btedit.bib} 

\end{document}